\documentclass[11pt]{JHEP3}

\usepackage{amssymb,amsmath}
\usepackage{graphics}
\usepackage{epsfig}
\usepackage{amsfonts}

\usepackage{amscd}

\def\be{\begin{eqnarray}}
\def\ee{\end{eqnarray}}

\def\g{\gamma}
\def\e{\epsilon}

\def\half{\frac{1}{2}}
\def\d{\partial}
\def\a{\alpha}

\def\vf{\varphi}

\newcommand{\idn}{{1\relax{\kern-.35em}1}}

\def\a{\alpha}

\def\g{\gamma}
\def\e{\epsilon}
\def\w{\omega}

\def\gg{\mathsf g }

\preprint{YITP-SB-09-36}

\title{From  Matrices to Strings and Back }
\author{Shlomo S. Razamat\footnote{Email: razamat@max2.physics.sunysb.edu}\\
C.N. Yang Institute for Theoretical Physics\\
State University of New York\\
Stony Brook, NY 11794-3840, U.S.A.}

\abstract{We discuss an explicit construction of a string dual for the Gaussian matrix model. Starting from
the matrix model and employing Strebel differentials techniques we deduce hints about the structure 
of the dual string. Next, following these hints a worldheet theory is constructed. The correlators
in this string theory are assumed to localize on a finite set of points in the moduli space of Riemann surfaces.
To each such point one associates a Feynman diagram contributing to the correlator in the dual matrix model,
and thus recasts the worldsheet expression as a sum over Feynman diagrams.}


\begin{document}

\section{Introduction}

One of the most fruitful directions of research in the recent years has been the exploration of the
connections between gravity theories in higher dimensions, string theory in particular, and gauge theories.

Although a significant number of gauge/string dualities is well established by now, we still are lacking
a ``microscopic'' understanding of the mechanisms behind the dualities in most cases. In cases when 
such mechanism is well understood we do not know how to apply it to more general setups. Examples of these
mechanisms are the Chern-Simons/topological string duality \cite{Gopakumar:1998ki,Ooguri:2002gx}, the Kontsevich model \cite{Witten:1990hr,Kontsevich:1992ti,Gaiotto:2003yb,Dijkgraaf:1991qh},
 and the double scaled matrix models/$c\leq1$ string theory duality \cite{Ginsparg:1993is,Klebanov:1991qa}.
See also~\cite{Berkovits:2007zk,Berkovits:2007rj,Berkovits:2008qc,Berkovits:2008ga} for a recent discussion of the $AdS_5/CFT_4$ duality.

The ``language'' naturally spoken by a perturbative gauge theory is that of sums of Feynman graphs. On the
other hand the ``language'' spoken by a string theory is that of integrals over moduli space of punctured Riemann surfaces.
The problem of having a microscopic understanding of a given instance of gauge/string duality ultimately boils down to finding a dictionary between
the two languages. That is, either figuring out how Feynman diagrams arise from a worldsheet
description of a string theory, or alternatively constructing a worldsheet from Feynman diagrams.
For instance, in the ``topological'' duality of  \cite{Gopakumar:1998ki,Ooguri:2002gx} the diagrams appear on the
string side as different classical solutions of the worldsheet theory. In case of double scaled matrix model \cite{Ginsparg:1993is,Klebanov:1991qa}
one obtains a string worldsheet by taking the large $N$ limit and tuning the couplings of the theory so that large Feynman
diagrams dominate and effectively can be treated as a discretization of a string worldsheet.

There exists a mathematical tool which can facilitate the translation between Feynman graphs and
moduli spaces of Riemann surfaces. Using a special class of quadratic differentials 
on a Riemann surface, the Strebel differentials, one can establish an isomorphism between the space of  metric graphs with a given number
of faces and the moduli space of punctured Riemann surfaces \cite{Strebel,Mulase:1998,Zvonkine:2002}.
 This tool has been used in different contexts in the framework of string theory, such as string field theory (see for example
\cite{Giddings:1986bp,Zwiebach:1992ie}) and matrix models (see \cite{Mukhi:2003sz} for a review).
Strebel differentials played an important role in understanding the Kontsevich model/$2d$ topological gravity duality \cite{Kontsevich:1992ti},
although by now we have additional ways to obtain microscopic understanding of this duality \cite{Okounkov:2000gx,Okounkov:2003gx}.

In~\cite{Gopakumar:2005fx} it was suggested  to use Strebel differentials to explicitly construct
 a generic class of gauge/string dualities in the free limit of the gauge theory side of the correspondence.
 See~\cite{Gopakumar:2003ns,Gopakumar:2004qb} for earlier work and 
 \cite{Furuuchi:2005qm,Aharony:2006th,David:2006qc,Aharony:2007fs,Yaakov:2006ce,David:2008iz} for subsequent developments.
The logic of this prescription is as follows. 
The isomorphism between moduli spaces and graphs implied by Strebel differentials needs a metric to be defined for the graphs. The choice
of metric effects the properties of the worldsheet dual. In \cite{Gopakumar:2005fx}  R.~Gopakumar introduced a metric on the Feynman graphs
through Schwinger parameters. One can write Feynman diagrams in a given field theory as integrals over Schwinger parameters.
 Then, using Strebel differentials this integration can be transformed to an integration over moduli space of punctured 
Riemann surfaces. The integrand of this integral is interpreted as a worldsheet correlator of a $2d$ $CFT$ dual of the field theory.
By computing enough worldsheet correlators in this way one hopes to gather enough evidence to deduce what is actually the string 
theory producing them.

Although the prescription of \cite{Gopakumar:2005fx} is very explicit it is technically hard to implement in practice.
Even in simplest cases making the transformation of measure between Schwinger and moduli 
parameters eventually reduces to solving elliptic integral equations which is usually impossible to do analytically.\footnote{
However, some correlators have been explicitly computed \cite{Aharony:2006th,David:2006qc,Aharony:2007fs}.} Thus, it is hard
to gather substantial amount of information from the correlators to completely specify the  string dual.

The main purpose of this paper is to discuss this program in a simplified setting. 
We make two simplifications. First, the field theory side is taken to be simply a (Gaussian) matrix model. Second, we change
the metric on the Feynman graphs.

 There is a natural metric on the Feynman graph which does not
require any additional structure to be introduced~\cite{Razamat:2008zr}. We can assign a unit of length to every edge of the diagram. 
Consider the following correlator in some gauge theory
\be\label{corr}
\langle\prod_{j=1}^sTrQ^{J_j}\rangle,
\ee where $Tr Q^{J}$ is a composite of $J$ adjoint fields in the theory. Then the set of all graphs dual to the connected diagrams
contributing to the calculation in the free limit is the set of  metric graphs with $s$ faces with circumferences of the
faces being $J_j$. Each diagram will be accompanied by an appropriate numerical factor computed as the number of the different Wick contractions 
giving this graph and appropriate momenta (position) dependence. Taking the gauge theory to be a matrix model 
one gets rid of the momentum (position) dependencies. 
 Using the Strebel theorem\footnote{We refer the reader to appendix~\ref{strebapp} or to~\cite{Gopakumar:2005fx,Aharony:2007fs,Razamat:2008zr} 
for brief expositions of the Strebel theorem.}
one can  map each Feynman graph
to a point in the moduli space of Riemann surfaces ${\mathcal M}_{\gg,s}$.  Thus, the correlator above
can be seen as a weighed sum over {\it discrete} set of points in the moduli space. However, in a usual string theory
the worldsheet correlators are given in terms of smooth integrals over the moduli space, and we will have to reconcile this issue.

In what follows we will construct a string dual of the Gaussian matrix model which will exhibit the properties discussed above.
The worldsheet theory will consist of three parts: a $c=1$ and a $c=-23$ CFTs which will be explicitly specified, 
and a $c=48$ CFT which will only be implicitly defined through some of its properties
. The correlators of the $c=48$ CFT will be {\it assumed} to localize the 
moduli space integration to a set of points corresponding to Strebel differentials with integer edge lengths.\footnote{For a
similar discussion in the case of a symmetric product orbifolds see \cite{PRR1}.}
We will refer to these special locations as {\it Strebel points}. As we will discuss in the bulk 
of the paper the Strebel points are in one to one correspondence with the Feynman diagrams contributing to a given correlator.
We will not be able to provide an explicit mechanism for such a localization. However,
assuming this property the sphere string theory correlators will be shown to reproduce planar matrix model results.

The paper is organized as follows. In section \ref{field} we review the results of \cite{Razamat:2008zr} and in particular starting
 from Gaussian matrix model
and using Strebel differential techniques gather some hints about its stringy dual. Following these hints in section \ref{string} we construct
a worldsheet model. In section \ref{proof} we show that certain correlators in this model reproduce the 
matrix model results. Finally, in section \ref{sum} we summarize  our results.
A short primer on Strebel differentials is included in appendix~\ref{strebapp}.
In appendix \ref{loc} we discuss a heuristic suggestion for the localization mechanism.

\section{From matrices to strings}\label{field}

In this section we discuss  the Gaussian matrix model and the hints it provides about its stringy  dual through
Strebel differentials.

\subsection{The matrix model}
We consider a gauge invariant correlator \eqref{corr}
 computed in the Gaussian matrix model with an action given by
\be\label{Maction}
S_M=\half\, t\,N\, Tr\,Q^2,
\ee where $Q$ is a hermitian $N\times N$ matrix.

 This correlator is computed as a sum of the different Wick contractions. It can be naturally expanded in powers of $N$.
 Every loop gives a factor of $N$,  and a propagator gives $(Nt)^{-1}$. All the  vertices  are external 
 and do not give any factors of $N$. Thus the power of $N$ accompanying  each diagram contributing to an $s$ point correlator \eqref{corr} is 
 $f-e=2-2\gg-s$ ($f$ is the number of loops and $e$ is the number of edges in a given diagram).  Thus, a   correlator \eqref{corr} in the free theory can
 be written as
\be\label{prefactor}
t^{-\half\sum_{j=1}^s J_s}\sum_{i} N^{2-2\gg_i-s}\,C_i= \prod_{k=1}^s J_k\,\prod_{J=1}^\infty v_J!\,N^{-s}\,t^{-\half\sum_{j=1}^s J_s}\sum_i \frac{N^{2-2\gg_i}}{\#(\Gamma_i)},
\ee
 where summation is over different Feynman graphs and $C_i$ is the number of Wick contractions giving each graph (for instance see \cite{Penner}). 
 $v_J$ is the number of vertices of power $J$. $\gg_i$ is the genus and the factor $\#(\Gamma_i)$ is the symmetry factor of
the $i$th Feynman graph (we identify the graphs by mapping the set of vertices and the set of edges onto themselves keeping the orientation of the vertices fixed).

\subsection{Extracting hints about the dual theory}
In this section we briefly review the construction of \cite{Razamat:2008zr}. The starting point is a variation of Gopakumar's 
prescription for constructing a string theory dual for a free field theory \cite{Gopakumar:2005fx}. In this variation 
we seek a string theory with the following property. For a given correlator the integration over the moduli space
of the Riemann surfaces localizes to a discrete set of points, with every point corresponding
to a Feynman diagram contributing to the calculation on the field theory side. Given a Feynman diagram
the corresponding point in the moduli space is computed through Strebel differentials by the following simple procedure.
We associate a unit of length to each edge of the Feynman diagram. Next, we make a \textit{skeleton} graph from the Feynman diagram \cite{Gopakumar:2005fx}
by gluing the homotopically equivalent edges together. The length of the edges of the skeleton graph is the number of edges of the Feynman graph
glued together. Then, the point in the moduli space corresponding to the Feynman diagram is the unique point corresponding
to the Strebel differential with the dual of the skeleton graph being its critical curve and the metric on the graph given by the 
lengths of the edges defined above.

The above construction by itself seems to teach us very little about the string theory side. However, one can extract some clues about the string theory
through the following procedure. Let us devise a ``caricature'' of string theory which will capture the Strebel differential construction
above and have some string theory structure to it.  
The correlators in the ``caricature'' model are computed using the following expression
\be\label{stringPv2}
\langle\prod_{k=1}^s{\mathcal O}_{J_k}\rangle_\gg&=&\left[\sum_{\vf\in{\mathcal S}_{\gg,s}}\frac{1}{\Gamma(\vf)}\right]\; 
\left[\prod_{k\in Z(\vf)\cup P(\vf)} \int_0^1dX_k\right]\,
 e^{-S}\, \prod_{k=1}^s{\mathcal O}_{J_k}.
\ee The first square brackets is a toy model for integration over moduli space. The set ${\mathcal S}_{\gg,s}$ is the set of all Strebel differentials,
$\vf$, on genus $\gg$ surface with $s$ punctures with integer edges.\footnote{
See e.g. \cite{Ashok:2006du,Alexandrov:2006qx,Chekhov:1993xp} for discussion of Strebel differentials with integer edges.}
 $\Gamma(\vf)$ is a discrete measure on the moduli space to be  specified shortly.
The second bracket gives a toy integration over a worldsheet ``field'' $X$. $X$ is defined only at the poles, $P(\vf)$, and zeros, $Z(\vf)$,
of the differential $\vf$. 
 The toy action, $S$, is given by
\be\label{action}
S&\equiv& 2\pi i\sum_{k\in P(\vf)\cup Z(\vf)} \left[p_k(\vf)-2\right] X_k 
-N\,\sum_{k\in Z(\vf)}\, e^{-4\pi i X_k},
\ee with $p(\vf)$ being the residues of the corresponding pole or zero of $\vf$. For a double pole it is the usual residue
of the square root of the differential and for a zero $p=0$.
 The operators are defined as\footnote{In this toy model we will not be able to incorporate the dilaton operator, $TrQ^2$.}
\be\label{opv2}
TrQ^J\,\to\;{\mathcal O}_J=N^{-J/2}
J\sum_{k\in P(\vf)} e^{2\pi i (J-2) X_k}.
\ee We specify $\Gamma(\vf)$ to be
\be\label{Gamma}
\Gamma(\vf)\;\equiv \; \#\Gamma(\vf) \times n(\vf),
\ee  where $\#\Gamma(\vf)$ is the symmetry factor of the critical curve of $\vf$ and $n(\vf)$ is the number of points on the moduli space having
the same graph as their critical curve.\footnote{Note that there is a unique Strebel differential for every graph with labeled vertices. However, in
the above construction we do not label the vertices and thus there are several points corresponding for a given graph.}

With the above definitions it is easy to convince oneself that the correlator computed on the matrix theory side,
$\langle\prod_{j=1}^sTrQ^{J_j}\rangle_\gg$ is exactly equal to the correlator $\langle\prod_{k=1}^s{\mathcal O}_{J_k}\rangle_\gg$
computed in the ``caricature'' string theory model. Strictly speaking the toy model gives the sum over connected
Feynman diagrams lacking homotopically trivial self contractions. If so desired, the homotopically trivial
self contractions can be easily taken into account by either redefining the set ${\mathcal S}_{\gg,s}$ or by redefining the operators
${\mathcal O}_J$ \cite{Razamat:2008zr}.  The integration over the fields $X_i$ does not vanish only for differentials
which have the set $J_i$ as the set of the residues of their double poles. Because we restrict ourselves to differentials with integer 
edges, all these differentials correspond to the Feynman diagrams contributing to \eqref{corr}. The integration can be easily performed and
the equivalence established.

Next, one can recast the toy model above into a bit more familiar language. The first term in the action \eqref{action}
can be written in the following form
\be
 2\pi i\sum_{k\in P(\vf)\cup Z(\vf)} \left[p_k(\vf)-2\right] X_k=-i\int d^2 z \sqrt{g_D}\,R_D\,X(z,\bar z),
\ee where the field $X(z,\bar z)$ is equal to $X_k$ at the special points of the differential and the metric $g_D$ is defined as follows. Given a Strebel differential $\vf$ with integer edges we can define
a \textit{dual} Strebel differential as (see section \ref{strebs} and~\cite{Razamat:2008zr}   for more details)
\be\label{vfD0}
\vf_D=-\frac{\vf}{\sin^2\pi l},\qquad l(z)=\int_{z_0}^z \sqrt{\vf}\, dz, 
\ee where $z_0$ is some (does not matter which) zero of $\vf$. The metric then is simply given by
\be
g_D=|\vf_D|. 
\ee
Moreover, the second term in \eqref{action} is heuristically of the following form
\be
 N\,\sum_{k\in Z(\vf)}\, e^{-4\pi i X_k}\;\to\;\mu\int d^2 z\,\sqrt{g_D}\, e^{-4\pi i X(z,\bar z)}.
\ee  Thus, the action can be schematically written as
\be\label{c25}
S\sim  -i\int d^2 z \sqrt{g_D}\,R_D\,X(z,\bar z)-\mu\int d^2 z\,\sqrt{g_D}\, e^{-4\pi i X(z,\bar z)}.
\ee Adding a kinetic term of time like boson to the above action we get a $c=25$ $CFT$. To see this we proceed as follows.
For the toy model to reproduce the matrix model results the ratio between the coefficients in the exponent and the coefficient
of the linear term has to be $4\pi$, as it is in \eqref{c25}. In the background charge representation of conformal field theories with
time like field,
\be
S=\frac{1}{4\pi}\int d^2 z\left[-\d\phi\bar\d\phi-\half\,i\, Q\sqrt{g}\,R\,\phi+\mu\,\sqrt{g}\,e^{ib\phi}\right],
\ee these coefficients are related due to conformal invariance by \be\label{Qb} Q=-\frac{2}{b}-b.\ee
Here $Q$ is the background charge and it is related to the central charge through $c=1+3Q^2$.
We can rescale the field $X$ by some constant $a^{-1}$ to obtain the relation \eqref{Qb}. We get the following equation
\be
8\pi a=\frac{2}{4\pi a}+4\pi a\quad\to\quad |a|=\frac{1}{2\sqrt{2}\pi} 
\ee
  Thus we deduce that in our case 
$Q=2\sqrt{2}$ and $c=25$. If one tries a space-like sign for the field $X$ then we again obtain $c=25$ but 
with the field $X$ rescaled by an imaginary number. As we have taken $X$ to be real in \eqref{c25} we are forced to consider $X$ to be time-like.  
 
The model above is a toy model without any obvious or direct relation to  a string theory construction. 
However, assuming that there is a string theory with properties described in the beginning of this
section one can try and extract some hints about this theory from the toy model above. 
 In the following sections we will try to construct
a string theory dual of the Gaussian matrix models starting from the following clues extracted from the toy model,
\begin{itemize}
 \item  The worldsheet theory contains a version of the $c=25$ $CFT$ obtained in \eqref{c25}.
 \item The metric $g_D$ might come handy in trying to prove that the string theory discussed is related to a matrix model.
 By definition a string theory is independent on a particular choice of the worldsheet metric due to\textit{Weyl} invariance so we
will have to understand what role the special metric $g_D$ might play.
\end{itemize}
We would like to stress that we  will not try to reproduce the toy model from a well defined string theory but rather 
follow the clues above to understand how to build a stringy dual for the Gaussian matrix model.  

\section{The string theory}\label{string}
We will now follow the hints of the previous section to construct a worldsheet dual for the Gaussian matrix model.
The string theory consists of three pieces: $\tilde X$ CFT with central charge $c=-23$, $Y$ CFT with $c=1$, and the $\chi$ CFT with
$c=48$. The logic is as follows. First, we will start with the worldsheet model motivated by the construction of the previous 
section. This model consists of a $c=25$ $\hat X$ CFT and a $c=1$ $\hat Y$ CFT. The $c=25$ $\hat X$ CFT is the one obtained from the toy
model construction of the previous section, and the $c=1$ $\hat Y$ CFT is introduced to render the total
 central charge of the matter fields to be $c=26$.
 After performing some elementary 
field redefinitions we will make a non trivial change of sign of one of the kinetic terms to recast the worldsheet
action as a sum of $c=-23$ $X$ CFT, $c=1$ $Y$ CFT, and a $c=48$ measure on the moduli space. We then interpret 
the $c=48$ measure as a result of computing a correlator of a $c=48$ $\chi$ CFT. 
Assuming that the moduli space integration localizes to the Strebel points, in the next section the worldsheet model  obtained in this fashion
is shown to reproduce planar Gaussian matrix model correlators.

\subsection{Constructing the action}
We start the construction from the following $c=26$ worldsheet theory.
The matter content consists of two $CFT$s, which we will denote as $\hat X$ and $\hat Y$.
The fields $\hat X$ and $\hat Y$ are taken to be periodic with period $1$.
The action is defined as
\be
S_{\hat X}(\hat \mu)&=&-2\pi\int d^2z\, \d \hat X\bar \d \hat X-i\int d^2 z\sqrt{g_B} R_B\, \hat X.\\
S_{\hat Y}(\hat \tau)&=&2\pi\int d^2z\, \d \hat Y\bar \d \hat Y\,.\nonumber
\ee  The $\hat X$ $CFT$ has central
charge $25$ and the $\hat Y$ $CFT$ has unit central charge.\footnote{
This worldsheet matter content was discussed by Itzhaki and McGreevy \cite{Itzhaki:2004te} as a candidate 
for a dual of large $N$ gauged harmonic oscillator \cite{Berenstein:2004kk}. One can transform 
to actions with usual kinetic terms by taking $\phi=-2\sqrt{2}\pi \hat X$.}
The metric $g_B$ is the ``dynamical'' metric with respect to which the measure of the path integral
is defined. In what follows we will assume that $g_B$ is in a conformal gauge and has the following form
\be
g_B(z,\bar z )\, dz\, d\bar z = |\vf_B(z)|\,dz\,d\bar z,
\ee where $\vf_B$ is a meromorphic function on the worldsheet. 
 Let us discuss this action in detail.
\subsection*{{\bf The $X$ $CFT$}}
The $\hat X$ $CFT$ has central charge $c=25$. One way to see this is by making a\textit{Weyl} transformation, $g_B\to e^{2\w}g_B$.
Under this transformation the field $\hat X$ transforms as 
\be \hat X\to \hat X-\frac{i}{\pi}\,\w,\qquad \sqrt{g_B}R_B\to \sqrt{g_B}R_B-4\d\bar\d\,\w.\ee
Thus we get
\be
\delta S_{\hat X}&=&-\frac{2}{\pi}\int d^2z\d\w\bar\d\w-\frac{1}{\pi}\int d^2z\sqrt{g_B} R_B\,\w=-24\,S_L(\w),
\ee where
\be
S_L(\w)=\frac{1}{12\pi}\int d^2z\left[\d\w\bar\d\w+\half\sqrt{g_B} R_B\,\w\right].
\ee
The measure of integration, $[{\mathcal D}\hat X]_{g_B}$ gives another factor of $S_L$ and thus we get that the $\hat X$ $CFT$ has $c=25$.

Note that we defined $\hat X$ to be a periodic scalar of period one, and this is in odds with the\textit{Weyl} transformation above. Thus, 
the field $\hat X$ should generally have some imaginary part. We make
 the following field redefinition
\be
  \hat X\equiv  X+\frac{i}{2\pi}\, \ln\left|\frac{\vf_D}{\vf_B}\right|\equiv X+\frac{i}{\pi}\hat\w.
\ee Here, $|\vf_D|\;(\,\equiv g_D\,)$ is some moduli dependent metric which does not transform under\textit{Weyl} transformations. This metric is related to the 
one discussed in the previous section and we will carefully define it below. The field $X$ is also\textit{Weyl} invariant and will be  the one integrated
over, i.e. $[{\mathcal D}\hat X]_{g_B}\to[{\mathcal D}X]_{g_B}$. Using this definition the $X$ field is periodic with period $1$ and has $c=25$.
One can think of the redefinition of $\hat  X$ merely as specifying the metric for which $\hat X$ has no imaginary part. This redefinition
will become non trivial shortly for two reasons. First, the operators will be defined through fields $X$ and not $\hat X$ and second
we will turn the field $X$ into a space-like field.
 We observe  the following,
\be
S_{\hat X}(0)=&&-2\pi\int d^2 z \d(X+\frac{i}{\pi}\hat\w)
\bar\d(X+\frac{i}{\pi}\hat\w)-i\int d^2 z\sqrt{g_B}R_B(X+\frac{i}{\pi}\hat\w)=\\
=&&-2\pi\int d^2 z\d X\bar \d X-i\int d^2z\sqrt{g_D}R_D\,X+\frac{2}{\pi}\left[\int d^2z\d\hat\w\bar\d\hat\w
+\half\int d^2z\sqrt{g_B}R_B\hat\w\right]\nonumber\\
=&&-2\pi\int d^2 z\d X\bar \d X-i\int d^2z\sqrt{g_D}R_D\,X+24\,S_L(\vf_B\to \vf_D),\nonumber
\ee where $S_L(\vf_B\to \vf_D)=S_L(\hat\w)$ is the Liouville action for going from metric $g_B$ to $g_D$.
Under\textit{Weyl} transformations  $g_D$ and $X$ are held fixed and $g_B$ transforming as usual. Thus, under\textit{Weyl} transformation
$g_B\to \exp(2\w')\, g_B$ the action transforms as
\be
\delta S_{\hat X}(0)=24\,S_L(e^{2\w'}\vf_B\to \vf_D)-24\,S_L(\vf_B\to \vf_D)=-24\,S_L(\vf_B\to e^{2\w'}\vf_B).
\ee 
Note that the field $X$ is time like and because it is \textit{Weyl} invariant, we can change the sign of the kinetic term
of $X$ (not of $\hat X$) without changing the central charge of the $CFT$. We will treat the field $X$ as spacelike and define
\be\label{xaction}
S_L&=&\,S_L(\vf_B\to \vf_D),\\
S_X(\hat \mu)&=&2\pi\int d^2 z\d X\bar \d X-i\int d^2z\sqrt{g_D}R_D\,X-\hat \mu {\mathcal O}_0.\nonumber
\ee The operator ${\mathcal O}_0$ will be defined in section \ref{opsec}.
Note that the theory $S_X+S_L$ is \textit{not} equivalent to $S_{\hat X}$ because of the change of the sign of the kinetic term.
 We merely used $\hat X$ action to construct the action of our interest. 
It is most natural to think of the $X$ $CFT$ as a $c=-23$ usual space-like theory with background charge accompanied by a measure factor with
\textit{Weyl} anomaly $c=48$. This comes about as follows.
We define 
\be 
S_X^{(\hat\w)}= S_X(\hat \mu=0)+24\,S_L(\hat\w), \qquad \hat\w=\half\ln\left|\frac{\vf_D}{\vf_B}\right|,
\ee and we can rewrite this as
\be
S_X^{(\hat\w)}=2\pi\int d^2 z\d X\bar\d X+4i\int d^2z \d\bar\d\hat\w X-i\int d^2z \sqrt{g_B}R_B \left(X+\frac{i}{\pi}\hat\w\right)+\frac{2}{\pi}\int d^2 z\d \hat\w\bar\d \hat\w\, .
\nonumber\\
\ee At this point $\hat\w$ can be a generic scalar function on the worldsheet transforming under \textit{Weyl} $g_B\to e^{2\w}g_B$ as
$\hat\w\to \hat\w- \w$. We can redefine the field $X$ by
\be
\tilde X\equiv  X-\frac{i}{\pi}\hat\w.
\ee Using this definition the action becomes
\be\label{xtildeaction} 
S_X^{(\hat \w)}&=&2\pi\int d^2 z\d \tilde X\bar\d \tilde X-i\int d^2z \sqrt{g_B}R_B \tilde X+\frac{4}{\pi}\int d^2z\left[\d\hat\w\bar\d\hat\w +
\frac{1}{2}\sqrt{g_B}R_B\hat\w\right],\nonumber\\
&=&S_{\tilde X}+48 S_L(\hat\w).
\ee Thus, we decompose the $c=25$ $X$ $CFT$  into a $c=-23$ $CFT$ $\tilde X$ and a measure which has \textit{Weyl} anomaly $c=48$.
Thus we can think of the $X$ action as $c=-23$ $CFT$ and a {\it Weyl} non invariant measure on the moduli space.
The $X$ action \eqref{xaction} is equivalent to the $\tilde X$ action \eqref{xtildeaction} and 
for computational reasons it will be more convenient to use the $X$ action in what follows.

\subsection*{{\bf The $Y$ $CFT$}}
For the $Y$ $CFT$ we make a trivial redefinition of the field,
\be
 \hat Y\equiv  Y-\frac{i}{2\pi}\,\ln|\Delta|,
\ee where $\Delta$ is some moduli dependent function on the worldsheet to be  specified later.
We obtain
\be
S_{\hat Y}(0)&&=2\pi\int d^2z \d(Y-\frac{i}{2\pi}\ln|\Delta|)\bar\d(Y-\frac{i}{2\pi}\ln|\Delta|)=\\&&=2\pi \int d^2 z\;\d Y\bar\d Y+
2i\int d^2z\left(\d\bar\d\ln|\Delta|\right) \, Y -\frac{1}{2\pi}\int d^2 z\;\d\ln|\Delta|\;\bar\d\ln|\Delta|.\nonumber
\ee
This is a $c=1$ $CFT$ as the action is independent on the metric. 
We will define
\be
S_\Delta&=&  -\frac{1}{2\pi}\int d^2 z\;\d\ln|\Delta|\;\bar\d\ln|\Delta|,\\
S_Y(\hat\tau)&=&2\pi \int d^2 z\;\d Y\bar\d Y+
2i\int d^2z\left(\d\bar\d\ln|\Delta|\right) \, Y-\hat\tau {\mathcal O}_{-2}.\nonumber
\ee The operator ${\mathcal O}_{-2}$
will be defined in section \ref{opsec}.
The above redefinition of the field $\hat Y$ will become non trivial later on, as the definition
of vertex operators  will depend on the fields $Y$. Thus, the explicit dependence on $\Delta$ 
contributes to the non standard measure on the moduli space.

\

\noindent To summarize, the matter action of our string theory is
\be\label{finaction}
\boxed{ S=S_X(\hat \mu)+S_Y(\hat \tau)+24\,S_L+S_\Delta}\;.
\ee
We regard this action as a combination of a $c=-23$ and $c=1$ CFTs together with a $c=48$ measure. We {\it assume}
that the $c=48$ part (with the measure dependent factors of section \ref{opsec}) can be obtained as correlators
in some $c=48$ CFT.

\

\subsection{Integration over the metric moduli}\label{moduli}
From now on we restrict our discussion to the planar case, i.e. the worldsheet is a sphere.
To compute the following correlator, 
\be
\langle\prod_{k=1}^s {\mathcal O_{J_k}}\rangle_{\gg=0},
\ee 
we have to fix the $SL(2,\mathbb{C})$ symmetry on the sphere. To do so we fix the positions of three of
the operators.
 The correlator has insertions with
$s$ quantum numbers $J_i$. For each point of the moduli space, i.e. for each set of positions of the unfixed operators, there is a unique Strebel
differential that has double poles at the positions of the operators with residues $J_k$. We denote this differential by $\vf$.
We define the correlator by the following expression
\be\label{intmod}
&&\langle\prod_{k=1}^s{\mathcal O_{J_k}}\rangle_{\gg=0}={\mathcal N}\,{\mathcal B}_{g_B}
\int_{{\mathcal M}_{0,s}}d\Omega\int\left[{\mathcal D}X{\mathcal D}Y\right]_{g_B}\, e^{-S}\,\prod_{k=1}^s{\mathcal O}_{J_k}\left(z_k(\Omega)\right),
\ee   where ${\mathcal N}$ is the
normalization factor and ${\mathcal B}_{g_B}$ are the ghost determinants, which for the sphere are just an overall factor. We parametrize the moduli space
by some coordinates $\Omega$.
The measure $d\Omega$ is taken to be \textit{Weyl} and  \textit{Diff} invariant. The action $S$ is defined in \eqref{finaction}. 
The complex number $z_k(\Omega)$ is the position of double pole  with residue $J_k$ at point $\Omega$ of moduli space.
Note that in this prescriptions ${\mathcal O}_J$ have to be $(0,\,0)$ operators. 

The set of points with all edges integer valued in $\vf$ metric correspond to Feynman diagrams.
Note, that each diagram will appear $n(\vf(\Omega))$ times. This is given by
\be
n(\vf)=\frac{1}{\#\Gamma(\vf)} \prod_{J=1}^\infty v_J!,
\ee  where $\#\Gamma(\vf)$ is the symmetry factor of the diagram and $v_J$ is the number of vertices of valence $J$.
One can understand the equality above as follows. Given  a diagram we have $\prod_{J=1}^\infty v_J!$ ways to assign labels 
to the vertices. However, the assignments can be grouped into equivalence classes with $\#\Gamma(\vf)$ 
elements in each class. The graphs are equivalent if there is a mapping between the two sets of  vertices and the two set of edges
which respects the ordering of the lines at each vertex. By Strebel theorem we have a unique point in the moduli space corresponding
for each class and thus there are $n(\vf)$ points corresponding to the diagram.

\subsection{Operators in the string theory}\label{opsec}
The operators we will discuss take the following form
\be\label{defops}
{\mathcal O}_{J}&=&\frac{1}{4\pi^2}\,J^3\,\left(\frac{\pi\hat \mu}{2}\right)^{-\half J}\,|\Delta|\;e^{2\pi i(|J|-2) X}\,e^{2\pi iJ Y}.
\ee The operators are \textit{not} normal ordered. The normalization is chosen for later convenience.
These are $(0,\,0)$ operators and can be used in the above described prescription.
  The two operators appearing in the
definition of the action, the ``puncture'' and the ``dilaton'' operators, take the following explicit form
\be
{\mathcal O}_0&=&\int d^2 z\,\sqrt{g_D}\;e^{-4\pi iX}\;,\qquad 
{\mathcal O}_{-2}=\int d^2 z\,\sqrt{g_D}\; e^{-4\pi i Y}\; .
\ee  Note that these operators are\textit{Weyl} invariant as the metric $g_D$ is the ``non-dynamical'' one.
Thus, there is no need to normal order the operators for this reason and we will see that 
there will be no need for renormalization in this prescription. In a sense the metric $g_D$ and the factors of $\Delta$ 
provide the renormalization needed for the correlators to be finite as we will see in what follows.
In what follows we will refer to the factors of $|\Delta|$ in the definition of ${\mathcal O}_J$, and to the factors
of $\sqrt{g_D}$ in ${\mathcal O}_{0,-2}$ as field independent terms.

\subsection{Definition of $g_D$ and $\Delta$ at the Strebel points}\label{strebs}

Let us specify the functions $g_D$ and $\Delta$ appearing in the definitions of the worldsheet theory. 
We will do so only for the Strebel points. For a heuristic discussion of a generic point in moduli space see appendix \ref{loc}.
To do so we make a short detour into properties of Strebel differentials with integer lengths.

\begin{figure}[htbp]
\begin{center}
\epsfig{file=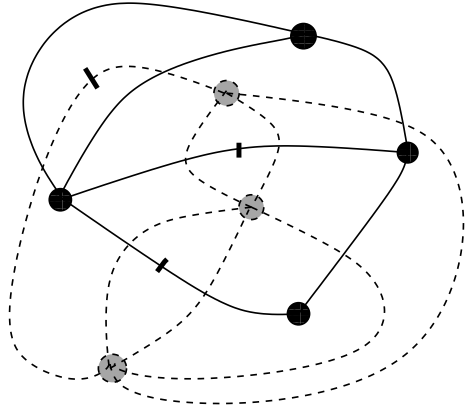,scale=0.33}
\caption{The solid line is an example of critical curve of $\vf$ and the dashed line is the critical curve of
corresponding $\vf_D$. The notches on the edges represent integer distanced points. One can observe that the two graphs 
are dual to each other.} \label{dualg}
\end{center}
\end{figure}

Given a specific correlator in the string theory 
\be\label{corr3}
\langle\prod_{k=1}^s{\mathcal O}_{J_k}\rangle_\gg,
\ee we define the metric $g_D$  as follows.
The set of all Strebel differentials with $s$ double poles is isomorphic to the decorated moduli space ${\mathcal M}_{\gg,s}\times {\mathbb R}_+^s$. Thus,
we parametrize the moduli space by the Strebel differentials which have the set $\{J_k\}$ as their residues. This is a  projection of the decorated 
moduli space to moduli space itself. The set of the Strebel differentials, $\vf$, satisfying the above property 
is isomorphic to ${\mathcal M}_{\gg,s}$. 
If all the edge lengths of  differential $\vf$ are integer valued we define (as in equation \eqref{vfD0})\footnote{Note that $l(z)$
has dimensions of length on the worldsheet. Thus, in equation \eqref{vfD} when $l(z)$ appears we essentially mean $l(z)/l_0$ with
$l_0=1$ for convenience. If so desired the factors of $l_0$ can reproduced in all the equations to follow.}
\be\label{vfD}
&&\vf_D\,dz^2\equiv -\frac{\,\vf}{\sin^2\pi l(z)}\,dz^2,\\
\tilde \vf\,dz^2&\equiv& -\frac{\,\vf}{\cos^2\pi l(z)}\,dz^2=-\frac{\,\vf_D}{\cos^2\pi l_D(z)}\,dz^2,\nonumber
\ee where we have used 
\be
l(z)\equiv \int_{z'}^{z}dz\sqrt{\vf},\qquad l_D(z)\equiv \int_{z_D'}^{z}dz\sqrt{\vf_D},
\ee and $z',\, z'_D$ are some given  zeros of the differential $\vf$ and $\vf_D$ respectively.  
As was shown in \cite{Razamat:2008zr} when all the edges of $\vf$ are integer the differential $\vf_D$ is essentially Strebel
(and the same proof holds for $\tilde\vf$).  The above three differentials satisfy
\be
\frac{1}{\vf}+\frac{1}{\vf_D}+\frac{1}{\tilde\vf}=0.
\ee
The critical curve of $\vf_D$ is the dual graph of critical curve of $\vf$.
For the sake of graph duality we regard integer distanced points as vertices (see figure \ref{dualg} for an illustration).
 The critical curve of $\vf_D$ at a Strebel point is one of the Feynman diagrams contributing to \eqref{corr3}.

 Because $l(z)$ is an integral of $\sqrt{\vf}$ we can translate the definition \eqref{vfD}
 into a first order non-linear differential equation for $\vf_D$,
 \be 
 4\pi^2\left(\vf_D+\vf\right)+\left(\d\ln\,\frac{\vf}{\vf_D}\right)^2=0.
 \ee
 This can be recast by defining $\vf_D=\vf\,e^{\hat\chi}$
\be \label{diffeqvfD0}
\left(\d\hat\chi\right)^2 + 4\pi^2\,e^{\hat\chi}\,\vf +4\pi^2\,\vf=0.
\ee Interestingly, the left hand side of the above equation  looks  like  a ``holomorphic Lagrangian'' for a chiral boson $\hat\chi$ with a Liouville interaction,
and with a meromorphic ``metric'' $\vf$.  We can integrate this equation for any $\vf$ to obtain
\be
e^{\hat\chi}=-\frac{1}{\sin^2\pi\left(l(z)+B\right)}, 
\ee where $B$ is some complex constant. For $B\in {\mathbb Z}$ we get $\vf_D$ and for $B\in {\mathbb Z}+\half$ we get $\tilde \vf$, and there
is a continuum of other boundary conditions. We can use \eqref{diffeqvfD0} to define $\vf_D$ for non-Strebel points, but then
the solution, $\hat\chi$, turns out to be not single valued and essentially is to be defined on some covering space of
the Riemann  surface. Thus, \eqref{diffeqvfD0} has well defined solutions only for Strebel differentials $\vf$ with integer edges.

We define metrics corresponding to the differentials,
\be
g=|\vf|,\qquad g'_D=|\vf_D|,\qquad \tilde g=|\tilde\vf|\,.
\ee Using
\be
g=e^{2\w}\;\to\; \sqrt{g}R=-4\,\d\bar\d\,\w\qquad (\,\sqrt{g_D}R_D=-2\,\d\bar\d\,\ln |\vf_D|\,)\,,
\ee
these metrics are associated with the following \textit{Ricci} scalars,
\be\label{curv}
-\frac{\sqrt{ g} R}{2\pi}&=&
\sum_{k\in P_D\cup  Z_D}m_k\delta^{(2)}(z-z_k),\qquad\\
-\frac{\sqrt{\tilde g}\tilde R}{2\pi}&=&\sum_{k\in Z_D}(p_k-2)\delta^{(2)}(z-z_k)
+\sum_{k\in P_D}m_k\delta^{(2)}(z-z_k)- 2\sum_{k\in E_D}\delta^{(2)}(z-z_k)\nonumber\\
-\frac{\sqrt{g'_D}R'_D}{2\pi}&=&\sum_{k\in Z_D\cup P_D}(p_k-2)\delta^{(2)}(z-z_k),\nonumber
\ee where the set $P_D$ is the set of double poles of $\vf_D$,
 the set $Z_D$ is the set of zeros and simple poles of $\vf_D$,
and the set $E_D$ is the set of edge centers of  $\vf_D$. We define the numbers $m_k$ to be the behavior of $\vf$ at a special point, i.e. 
$\vf\sim z^m$. The numbers $p_k-2$ are the behavior of $\vf_D$ at a special point, i.e. $\vf_D\sim z^{p-2}$.
From this we deduce that
\be\label{intercorr}
 &&\frac{\sqrt{g}R - \sqrt{\tilde g}\tilde R}{4}= \d\bar\d\,\ln |\tan \pi l_D|= \pi \left[ \half \sum_{k\in Z_D} p_k\delta^2(z-z_k)
-\sum_{k\in E_D}\delta^2(z-z_k)\right]\, ,\nonumber\\
&&\d\bar\d\,\ln |\vf_D|= \pi \sum_{k\in Z_D\cup P_D} (p_k-2)\delta^2(z-z_k)\,.
\ee 
We summarize the  behavior of the differentials in vicinity of special points in the following table,
\be\label{tablebeh}
\begin{tabular}{|c|c|c|c|}
\hline
&$\vf_D$ & $\tan^2\pi l_D$&$\vf$\\
\hline\hline
$z\in Z_D$&$Cz^{p-2}$ & $\frac{4\pi^2}{p^2}C\,z^p$&$-\frac{p^2}{4\pi^2}\frac{1}{z^2}$\\\hline
$z\in P_D$&$-\frac{(m+2)^2}{4\pi^2}\frac{1}{z^2}$ & $1$ & $B\,z^m$\\\hline
$z\in E_D$&$C$ & $\frac{1}{\pi^2C}\frac{1}{z^2}$&$-C$\\\hline
\end{tabular}\;,
\ee  where $B$ and $C$ are some complex numbers.

\

\noindent We are  finally ready to define the quantities $g_D$ and $\Delta$ at the Strebel points,
\be
 &&g_D=\frac{g'_D}{g'_D(\infty)}=
\left|\frac{\prod_{k\in Z_D}(z-z_k)^{p_k-2}}{\prod_{l\in P_D}(z-z_l)^2}\right|\,,\label{vfDinf}\\ 
&&\Delta=-\frac{1}{\cos^2\pi l}=\tan^2\pi l_D,\label{Delta}
\ee where we also assumed for simplicity that $\infty\notin P_D$.

\section{From strings to matrices}\label{proof}

In this section we will explicitly show how the worldsheet model introduced in the previous section
reproduces the matrix model results.
We remind that the explicit discussion is restricted to the planar topology of the worldsheet.
The exact claim is that contributions from planar and connected matrix model diagrams lacking homotopically trivial
self contractions to the following correlator,
\be
\langle\prod_{j=1}^sTrQ^{J_j}\rangle_{\gg=0},
\ee are reproduced by the worldsheet correlator 
\be\label{stringside}
\langle\prod_{j=1}^s{\mathcal O}_{J_j}\rangle_{\gg=0}\equiv \int_{{\mathcal M}_{0,s}} d\Omega\; {\mathcal I}.
\ee Following the procedure of section \ref{moduli} we insert  the operators ${\mathcal O}_{J_j}$
at the double poles of $\vf$ (denoted in what follows by $z_k$).
We denote the path integrals of the fields $X$ and $Y$ as,
\be
{\mathcal I}_X&\equiv&\int \left[ {\mathcal D} X\right]_{g_B} e^{-S_X}\prod_{k=1}^s\,e^{2\pi i(J_k-2)\, X(z_k)},\\
{\mathcal I}_Y&\equiv&\int \left[ {\mathcal D} Y\right]_{g_B} e^{-S_Y}\prod_{k=1}^s\,e^{2\pi i\, J_k\, Y(z_k)}.\nonumber
\ee We thus discuss these quantities at the Strebel points and assume that the integration over the moduli space localizes on 
Strebel points.

\

 The integration over the zero mode for the $X$ field implies that in the expansion
of the exponential of the interaction ${\mathcal O}_0$ only the product of $f$ interaction terms contributes. Here $f$ is the number of the faces
 of the critical curve of $\vf_D$, i.e. the number of faces of the Feynman diagram corresponding
to the particular Strebel point. The zero mode of the $Y$ field
fixes the number of the interaction insertions ${\mathcal O}_{-2}$ to be $e$, with $e=\half\sum_{k=1}^sJ_k$, the number of 
edges of the diagram. For the above to hold it is essential that 
both fields $X$ and $Y$ are periodic with period $1$, i.e. the zero modes take value in $[0,1]$.
Using \eqref{curv} to compute the linear terms in the $X$ and $Y$ actions we obtain the following
\be\label{corrs1}
{\mathcal I}_X&=&({\det}'\Delta_{g_B})^{-1/2}\,\frac{\hat \mu^f}{f!}\int \prod_{k=1}^f\left[d^2 z'_k\,\sqrt{g_D(z'_k)} \right]
\prod_{m,n=1}^f\frac{|z_{mn}|^2|z'_{mn}|^2}{|z_{m}-z'_n|^4},\\
{\mathcal I}_Y&=&({\det}'\Delta_{g_B})^{-1/2}\,\frac{\hat \tau^{e}}{e!}
\int \prod_{k=1}^e\left[d^2 \tilde z_k\,\sqrt{g_D(\tilde z_k)}\right]
\prod_{m,n=1}^e\frac{|\hat z_{mn}|^2|\tilde z_{mn}|^2}{|\hat z_{m}-\tilde z_n|^4},\nonumber
\ee where $z_k$ are the locations of double poles of $\vf_D$, $\hat z_k$ the locations of edge centers of $\vf_D$, $z'_k$ 
are the locations of ${\mathcal O}_0$, and $\tilde z_k$ are locations of ${\mathcal O}_{-2}$ interactions.
We have to integrate the above expressions over the interaction insertions $z'$ and $\tilde z$.
These integrals have to be regularized, and we do so by regularizing the expressions for the Green's function needed 
to compute \eqref{corrs1} as
\be
G(z,z')\sim-\ln(|z-z'|^2+\e^2),
\ee where $\e$ is a small real number which will be sent to zero at the end. We also regularize the differential $\vf_D$ in the vicinity of its zeros and poles as 
\be |\vf_D|\sim C_k\, (z\bar z+\e^2)^{\half(p_k-2)}.\ee Moreover note that 
\be
\delta^{(2)}_{\e,k}(z,\bar z)=\frac{k-1}{\pi}\frac{\e^{2(k-1)}}{(|z|^2+\e^2)^{k}},
\ee is a representation of the $\delta$-function on the complex plane for any $k> 1$ in the limit of small $\e$.

\subsection*{The $X$ $CFT$ part}

First, we deal with the $X$ part of the correlator. With the above regularization we can write,
\be \label{IX}
{\mathcal I}_X&=&({\det}'\Delta_{g_B})^{-1/2}\,\frac{(\hat \mu/\e^4)^{f}}{f!}
\int \prod_{l=1}^fd^2z'_l\,\e^{4f}\prod_{m,n=1}^f\frac{\left(|z_{mn}|^2+\e^2\right)\left(| z'_{mn}|^2+\e^2\right)}{\left(|z_{m}- z'_n|^2+\e^2\right)^3}\dots\nonumber\\
&=&\left(\frac{\pi\hat \mu}{2}\right)^{f}\,({\det}'\Delta_{g_B})^{-1/2}\;\prod_{k\in P_D} |C_k|.
\ee Here $C_k$ is the residue of $\vf_D$ at pole $z_k\in P_D$. We used the $\delta$-function representation above with $k=3$, and take the $\e\to 0$ limit. The extra power 
of $\left(|z_{m}- z'_n|^2+\e^2\right)$ comes from the $g_D$ factor in the definition of the puncture operator. The dots
in the first line represent the regular terms coming from  $g_D$ factor, see \eqref{vfDinf}.
 Thus we  get a finite result. The insertions of ${\mathcal O}_{0}$ are localized at the face centers. 

\subsection*{The $Y$ $CFT$ part}
Next, we deal with the $Y$ part of the correlator. With the above regularization we can write at the Strebel
points,
\be \label{IY}
{\mathcal I}_Y&=&({\det}'\Delta_{g_B})^{-1/2}\,\frac{(\hat \tau/\e^2)^{e}}{e!}
\int \prod_{l=1}^ed^2\tilde z_l\,\e^{2e}\prod_{m,n=1}^e\frac{\left(|\hat z_{mn}|^2+\e^2\right)\left(|\tilde z_{mn}|^2+\e^2\right)}{\left(|\hat z_{m}-\tilde z_n|^2+\e^2\right)^2}\dots\nonumber\\
&=&\left(\pi\hat \tau\right)^{e}\,({\det}'\Delta_{g_B})^{-1/2}\;\e^{2e}\;\;\prod_{k\in E_D} |C_k|.
\ee Here $C_k$ is the value of $\vf_D$ at edge center $z_k\in E_D$.  As we will see shortly there is  an additional factor of $\e^{-2e}$ coming from the field independent terms
which will cancell $\e^{2e}$ term above. We will get a finite result.
The insertions of ${\mathcal O}_{-2}$ are localized at the edge centers. 

\subsection*{Field independent terms}
Let us evaluate the actions $S_L$ and $S_\Delta$ and  the
field independent factors in the definition  of the operators at the Strebel points.

We compute the Liouville action $S_L(g_B\to g_D)$.
 For concreteness we choose a simple gauge on the sphere with the  metric given by
\be
\vf_B=-\frac{\a^2}{z^2},\qquad g_B=|\vf_B|.
\ee Because the model is {\it Weyl} invariant the result will not dependend on the detais of this metric.
The Liouville factor is
\be
\w=\half \ln |\vf_D|+\half \ln |z|^2-\half\ln|\a|^2,
\ee
and the Liouville action is
\be
 S_L&=&\frac{1}{12\pi}\int d^2z\left[\d\w\bar\d\w+\half \sqrt{g_B} R_B\,\w\right].
\ee
Explicitly we get
\be\label{SL1}
 S_L&=&-\frac{1}{48\pi}\int d^2z\biggl[\ln|\vf_D|\bar\d\d\ln|\vf_D|+2\ln|\a|^2\bar\d\d\ln|z|^2-\\
&&-\ln|z|^2\bar\d\d\ln|z|^2+\ln|z|^2\bar\d\d\ln|\vf_D|-\bar\d\d\ln|z|^2\ln|\vf_D|\biggr].\nonumber
\ee
The first term is given by
\be
-\frac{1}{48\pi}\int d^2z \,\ln|\vf_D|\d\bar \d\ln|\vf_D|=-\frac{1}{48}\sum_{z_k\in P_D\cup Z_D}
\left(\ln |C_k| +(p_k-2)\ln\e\right)(p_k-2).
\ee   The number $\e$ is some real, small regulator and in the end
we will take it to zero. Constants $C_k$ are the residues of $\vf_D$ at the special points, see table \ref{tablebeh}.
 Two last terms in \eqref{SL1} sum up to
\be
&&\int d^2z\,\ln|z|^2\bar\d\d\ln|\vf_D|=2\pi (\ln|\vf_D(0)|-\ln|\vf_D(\infty)|),\\
&&\int d^2z\,\bar\d\d\ln|z|^2\ln|\vf_D|=2\pi(\ln|\vf_D(0)|+\ln|\vf_D(\infty)|),\nonumber\\
&&\to\qquad-\frac{1}{48\pi}\int d^2z\biggl[\ln|z|^2\bar\d\d\ln|\vf_D|-\bar\d\d\ln|z|^2\ln|\vf_D|\biggr]=\frac{1}{12}\ln|\vf_D(\infty)|.\nonumber
\ee Remember that we have set $\vf_D(\infty)=1$ (see \eqref{vfDinf}) and thus the above vanishes.
 The last term we did not discuss gives
\be
\frac{1}{48\pi}\int d^2z\,\left(\ln|z|^2\bar\d\d\ln|z|^2-2\ln|\a|^2\bar\d\d\ln|z|^2\right)=\frac{1}{6}\ln \e-\frac{1}{6}\ln|\a|^2.
\ee Combining all the factors together we get
\be\label{liou}
-24\,S_L= 4\ln|\a|^2-4\ln\e +\frac{1}{2}\sum_{z_k\in P_D\cup Z_D}
\left(\ln |C_k| +(p_k-2)\ln\e\right)(p_k-2).
\ee

\noindent From the $S_\Delta$ action we get the following for Strebel points
\be\label{Delac}
S_\Delta&=& -\frac{2}{\pi}\int d^2 z\, \d\ln|\tan\pi l_D|\bar \d\ln|\tan\pi l_D|=\\
&=&\sum_{z_k\in Z_D}\,\left(\half\ln|C_k|+\ln\frac{2\pi}{p_k}+\half p_k\ln\e \right)p_k+
2\sum_{z_k\in E_D}\,\left(\half\ln|C_k|+\ln\pi+\ln\e \right).\nonumber
\ee

\

Using the above we compute the finite contributions from the field  independent terms, which we will denote by $\exp\left({\mathcal A}\right)$.
The Liouville action, $S_L$ \eqref{liou}, gives the following contribution 
\be
{\mathcal A}_L=\half \sum_{z_k\in Z_D\cup P_D}(p_k-2)\ln |C_k|,
\ee The $S_\Delta$ action \eqref{Delac} contributes
\be 
{\mathcal A}_{\Delta}= -\half\sum_{z_k\in  Z_D}p_k\ln |C_k|-\sum_{z_k\in E_D}\ln |C_k|.
\ee
For the external operators we get from \eqref{defops} using \eqref{tablebeh}
\be
{\mathcal A}_{\mathcal O_J}=\sum_{z_k\in  Z_D}\ln |C_k|,\qquad
\ee   and from the interactions we obtain (using the results \eqref{IX} and \eqref{IY})
\be
{\mathcal A}_{\mathcal O_0}=\sum_{z_k\in  P_D}\ln |C_k|,\qquad
{\mathcal A}_{\mathcal O_{-2}}=\sum_{z_k\in E_D}\ln |C_k|.
\ee
Thus we note that the finite field independent terms cancel out at the Strebel points,
\be
{\mathcal A}={\mathcal A}_L+{\mathcal A}_\Delta + \sum_{k\in Z_D} {\mathcal A}_{\mathcal O_{J_k}}+\sum_{k\in P_D}^f {\mathcal A}_{\mathcal O_0}
+\sum_{k\in E_D}^e {\mathcal A}_{\mathcal O_{-2}}=0.
\ee Next, the $\e$ dependence from the field independent terms is as follows,
\be
&&S_L\to\half \sum_{z_k\in Z_D\cup P_D}(p_k-2)^2\ln \e-4\ln\e,\qquad
S_\Delta\to	-\half \sum_{z_k\in Z_D}p_k^2\ln \e-2\sum_{z_k\in E_D}\ln \e,\nonumber\\
&&{\mathcal O_J}\to \sum_{z_k\in Z_D}p_k\,\ln\e,\quad {\mathcal O_0}\to -2\sum_{z_k\in P_D}\ln\e
,\quad {\mathcal O_{-2}}\to 0.
\ee These sum up to \be\label{epsadd} -4\ln \e-4e\,\ln\e+2s\,\ln\e=(-2e-2f)\,\ln\e.\ee  The factor $-2f\ln\e$ has been already taken into account
in the $X$ $CFT$ calculation, and the factor $-2e\ln\e$ cancels the $2e\ln\e$ contribution obtained in $Y$ $CFT$ calculation.

\

\noindent To summarize, at Strebel points the calculation is finite and is precisely given by 
\be
\left.{\mathcal I}\right|_{{\text{\tiny{\it Strebel}}}}={\mathcal N}{\mathcal B}_{g_B}\,|\a|^8\,({\det}'\Delta_{g_B})^{-1}\,\left(\frac{\pi\hat \mu}{2}\right)^{f}\,\left(\pi\hat \tau\right)^{e} \,,
\ee where ${\mathcal I}$ is defined in \eqref{stringside}. Collecting all the terms and assuming that the only contributions
to the moduli space integration  come from the Strebel points we get
\be
\langle\prod_{k=1}^s{\mathcal O}_{J_k}\rangle_{\gg=0}={\mathcal N}{\mathcal B}_{g_B}|\a|^8({\det}'\Delta_{g_B})^{-1}
\,\left(\frac{\pi\hat \mu}{2}\right)^{f-e}\, \left(\pi\hat \tau\right)^{e}
 \prod_{k=1}^s J_k\prod_{J=1}^\infty v_{J}!
\sum_i\frac{1}{\#\Gamma_i} 
\ee Identifying $\hat\mu=2N/\pi$, $\hat\tau=1/\pi\,t$, and choosing normalization\footnote{Note that 
essentially ${\mathcal N}$ is independent of $\a$
due to  {\it Weyl} invariance.}
 \be {\mathcal N}=\frac{{\det}'\Delta_{g_B}	}{{\mathcal B}_{g_B}|\a|^8},\ee we get
\be 
\langle\prod_{k=1}^s{\mathcal O}_{J_k}\rangle_{\gg=0}=\langle\prod_{k=1}^sTrQ^{J_k}\rangle_{\gg=0}.
\ee

\

\begin{figure}[t]
\begin{center}
\epsfig{file=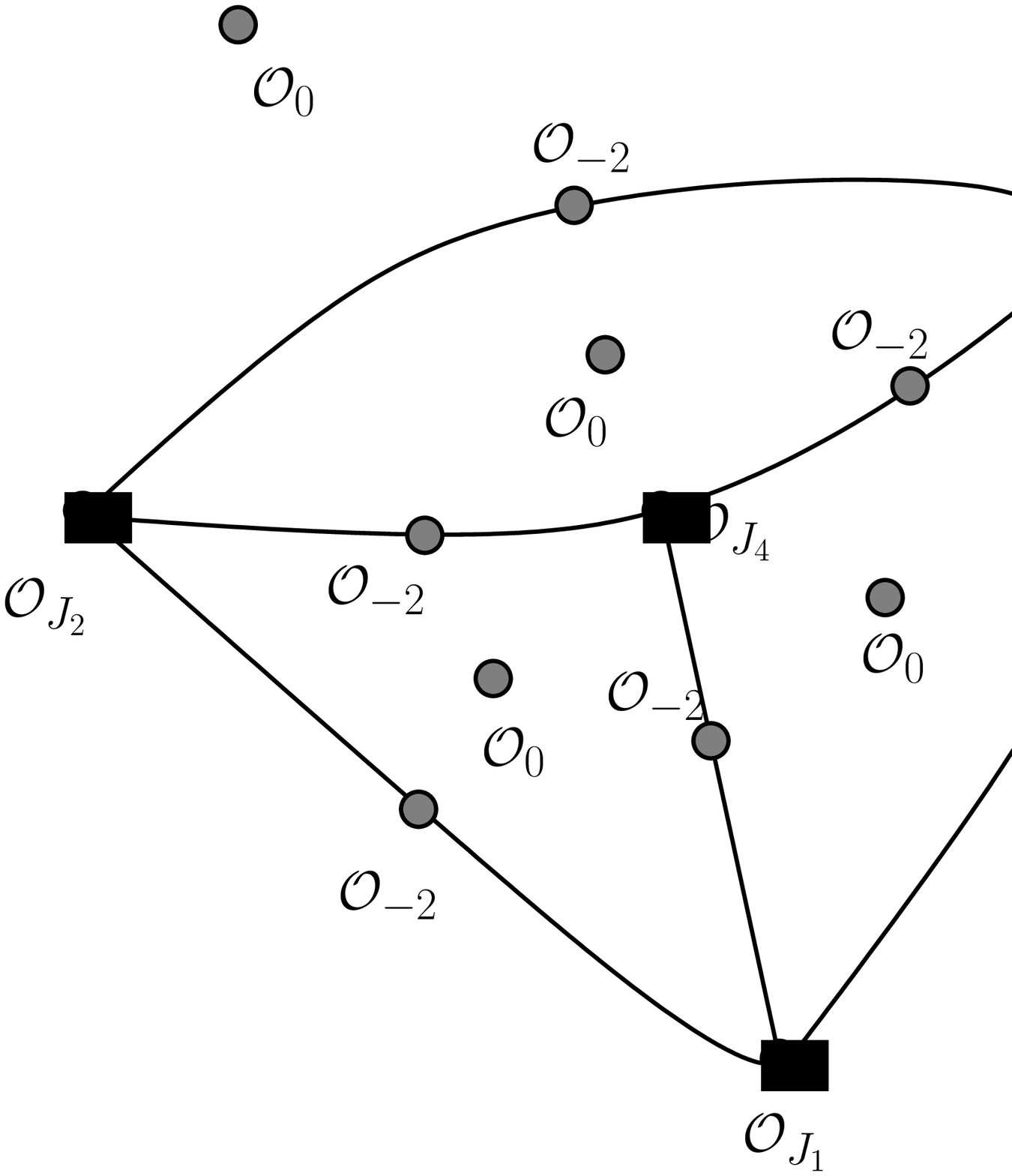,scale=0.33}
\caption{An example of a diagram with the interaction vertices localized at edge and face centers.} \label{findiag}
\end{center}
\end{figure}

\section{Summary}\label{sum}

Let us briefly summarize  our results.
We have defined the following worldsheet model.
 The action  is given by
\be
&c=-23 \quad&:\quad S_{\tilde X} = 2\pi\int d^2 z\d \tilde X\bar\d \tilde X-i\int d^2z \sqrt{g_B}R_B \tilde X-\hat \mu\, {\mathcal O}_0,\\
&c=1 \quad&:\quad S_{\hat Y}=2\pi\int d^2 z\d \hat Y\bar\d \hat Y-\hat\tau\, {\mathcal O}_{-2},\\ &c=48 \quad&:\quad S_{\chi}.
\ee The integrated operators are defined as
\be
{\mathcal O}_{J}=\int d^2 z \sqrt{g_B} \,e^{2\pi i (|J|-2)\,\tilde X}\,e^{2\pi i \,J\,\hat Y}\,{\mathcal O}^{(\chi)}_{J}\,.
\ee
We do not explicitly know the action $S_\chi$ and the form of ${\mathcal O}^{(\chi)}_{J}$.
However, assuming that this objects have certain properties we have shown that sphere correlators in the above string theory reproduce
planar Gaussian matrix model correlators. The assumptions on the field $\chi$ are as follows. First, we assume that the path integral
over $\chi$ localizes the moduli space integration to a discrete set of points,\footnote{In appendix \ref{loc} we present a
heuristic suggestion for the actual localization mechanism.} the Strebel points. Next, 
this path integral takes an explicit form at the Strebel points. In the language of the bulk of the paper 
this is responsible for the ``field independent'' terms in the calculation of the $X$ and $Y$ path integrals.

For the construction of this paper to be complete it will be very crucial to understand whether there actually is a $c=48$ $CFT$ 
with the above structure of correlators. As these correlators presumably depend on $\vf_D$,
this $CFT$ is expected to be defined through Strebel differentials or give rise to these differentials in some manner.\footnote{
For instance, in the context of closed string field theory the Strebel differentials appear as giving rise to a minimal
area metric \cite{Zwiebach:1992ie}.} We leave the investigation of these issues for future research.

\acknowledgments 

I am grateful to O.~Aharony, R.~Gopakumar, A.~Pakman, and L.~Rastelli
 for very useful discussions and comments on the manuscript. I would like to thank the organizers 
of the Monsoon Workshop on String Theory and the HET Group at the Weizmann Institute for hospitality
during different stages of this project. This work is supported in part by DOE grant DEFG-0292-ER40697
and by NSF grant PHY-0653351-001. Any
opinions, findings, and  conclusions or recommendations expressed in this 
material are those of the author and do not necessarily reflect the views of the National 
Science Foundation.

\appendix 

\section{A short primer on Strebel differentials}\label{strebapp}

A quadratic differential is the following object,
\be
q=\vf(z)\, dz^2,
\ee where $\vf$ is a meromorphic function on a given Riemann surface. This differential is defined to have the following property under a holomorphic reparametrization of the worldsheet
 $z\to z'(z)$,
\be
\vf(z)dz^2=\vf'(z')(dz')^2.
\ee Using quadratic differentials one can define a length for a line element through
\be\label{strmet}
dl=\sqrt{\vf}\,dz.
\ee Note that this length is in general a complex number.
It is useful to define the notions of horizontal and vertical curves of the differential.
Given a curve $\g(t)$ on the Riemann surface we say that it is horizontal if 
\be
\vf(\g(t))\left(\frac{d\g}{d t}\right)^2>0,
\ee and vertical if the opposite inequality holds. Note that the length
of the horizontal curves computed using \eqref{strmet} is real. By convention we will discuss 
the horizontal curves in what follows. A horizontal curve can either be closed or
end on a zero or a pole.  
The set of all non-closed horizontal curves of a quadratic differential is called the critical
curve of the differential. We restrict to quadratic differentials critical curve
of which is compact.
If a quadratic differential has at most double poles (with negative coefficients)
 then the critical curve divides the Riemann surface into ring domains. The vertices of the critical curve of such a differential are the zeros and the simple poles of the differential.
The following theorem due to K.~Strebel holds,

\noindent\textit{
Given a Riemann surface with $s$ marked points and $s$ positive numbers $p_k$ associated
to those points, there is a unique quadratic differential with double poles as its only
singularities such that:
\begin{itemize}
\item It has exactly $s$ double poles located at the marked points
\item The residues of the double poles are the numbers $p_k$
\item The Riemann surface is a union of $s$ disc domains defined by the marked points. 
\end{itemize}
}

We refer to a differential which satisfies the properties above as a Strebel differential.
Note that from this theorem follows that there is a unique  Strebel differential for each point of 
 ${\mathcal M}_{\gg,s}\times{\mathbb R}_+^s$. Further, this also gives us a natural
isomorphism between the space ${\mathcal M}_{\gg,s}\times{\mathbb R}_+^s$ and the space
of metric graphs with $s$ faces on genus $\gg$ surface.\footnote{In this context we define a metric graph as a connected graph with
a positive real number associated to every edge and all the vertices at least trivalent.} For each point of
${\mathcal M}_{\gg,s}\times{\mathbb R}_+^s$ we associate the critical curve of the corresponding
Strebel differential as the metric graph (metric on the graph defined through \eqref{strmet}),
and the other direction of the isomorphism can be also (less trivially) established. 

When explicitly trying to find a Strebel differential for a given Riemann surface and a given set
 of residues the first two conditions above can be easily satisfied. The third condition is
however a very non-trivial one. Essentially, it can be rephrased as the requirement that all the 
distances between the zeros of the differential computed in Strebel metric \eqref{strmet} should
 be real. Computing these distances will give constraints on the parameters of the differential. 
 Usually these constraints will be expressed through elliptic integrals, which are difficult to solve.

\section{A localization mechanism}\label{loc}
In this section we discuss a possible definition of the quantities $g_D$ and $\Delta$ for arbitrary points of the moduli space.
The definition of these quantities is closely related to the mechanism responsible for the moduli space integration localizing
on Strebel points. 

As was discussed in the bulk of the paper for a given $s$ point correlator and a given point of the moduli space one can associate
a Strebel differential $\vf$.
 If not all the edges of $\vf$ are integer valued some of the zeros of $\vf$ are non-integer distanced.
 In this case \eqref{vfD} is not well defined 
and we have to refine it. Demanding that for Strebel points $g_D$ is defined as in \eqref{vfD} we can have several possible definitions of
$\vf_D$ at a generic point in the moduli space. We will seek a definition which has a similar structural form, i.e. $\vf_D$ is a product of
the differential $\vf$ and another function defined in terms of some lengths computed using $\vf$. 

Note that at Strebel points the nature of poles and zeros of $\vf_D$ as defined in \eqref{vfD} is very different. 
The zeros of $\vf_D$ come from divergent lengths, $l(z)$, near a pole of $\vf$.  The pole combines with the divergence
of the edge length and we get a zero. Thus in an extension structurally
similar to \eqref{vfD} this will be usually also the case.  On the other hand 
near a zero of $\vf$ we get a pole at the Strebel points \textit{only} because the position of the zero coincides with the point at which $l$ is an integer.
This fact is very special to the Strebel points. Thus, the structure of the poles and zeros of $\vf_D$ will 
 change near the zeros of $\vf$ for a generic refinement of \eqref{vfD}.

Let us consider a simple refinement of the definition of $\vf_D$ which we can only directly define
in a vicinity of the zeros of $\vf$. The special feature of this extension is that it is explicitly tractable. Although we do not know how to extend this definition over the whole moduli space it has qualitative features which are robust, i.e. appearance of extra zeros and poles near Strebel points.
Of course it would be very interesting to find a definition
of $\vf_D$ valid for the entire worldsheet and even more interesting to come
 up with an argument which will favor a certain definition of $\vf_D$.

\begin{figure}[htbp]
\begin{center}
$\begin{array}{c@{\hspace{0.45in}}c}
\epsfig{file=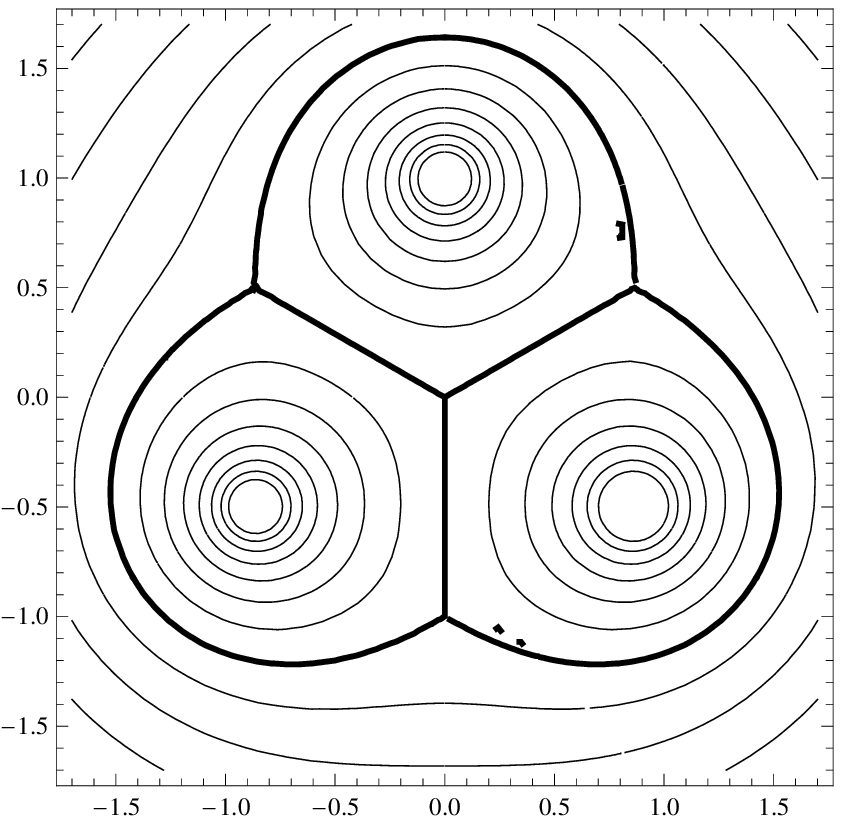,scale=0.75} & \epsfig{file=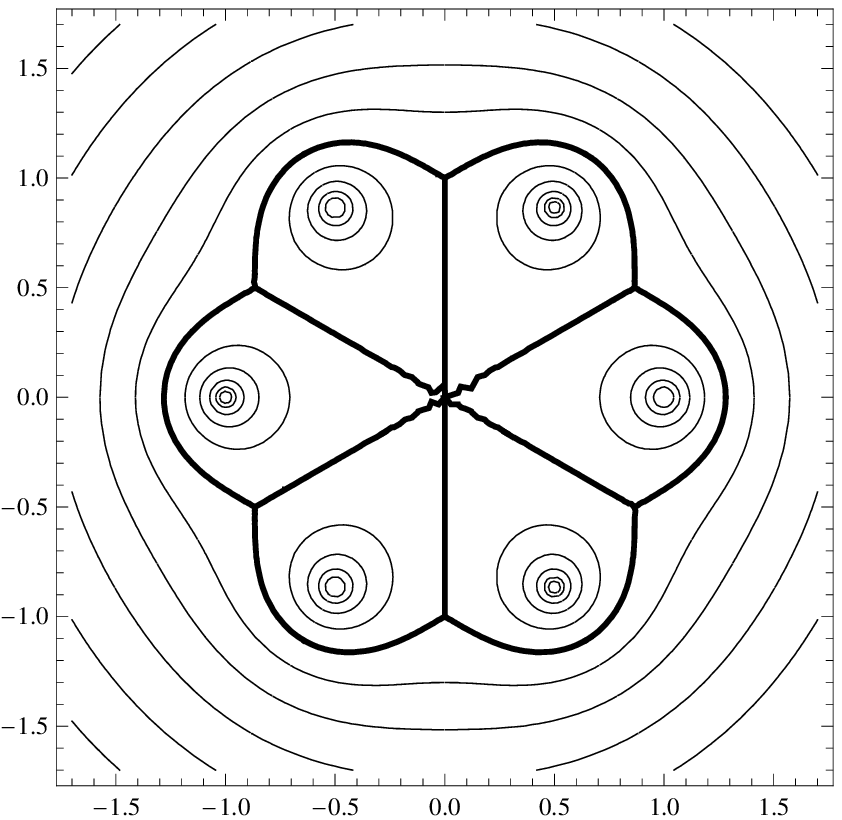,scale=0.75}\\ [0.2cm]
\end{array}$
\caption{On the left we have $ \vf^{(m=1)}$ with ${\hat z}^{m+2}=i$, and on the right $ \vf^{(m=4)}$ with ${\hat z}^{m+2}=1$.
The edges emanating from $z=0$ have length $\half\left(\sqrt{2}-1\right)$, and the outer edges are all unit length 
(this holds for any $m$ and any $\hat z$ ).} \label{strebFace} 
\end{center}
\end{figure}

Near a Strebel point ( and a zero of $\vf$) we define $\vf_D$ in the following way
\be\label{faceStreb}
\vf_D\sim-\frac{(m+2)^2}{4\pi^2}z^m\frac{\hat z^{m+2}+z^{m+2}}{\left(\hat z^{m+2}-z^{m+2}\right)^2}\equiv \vf^{(m)}, 
\ee where 
 \be \hat z\sim \left(\delta^2\,C\right)^{1/(m+2)}\,.\ee
The parameter $\delta$ is the fractional part of the distance of the zero around which we define the differential to some reference zero.
Taking $\delta\to 0$ the differential $\vf^{(m)}$ just becomes $-\frac{(m+2)^2}{4\pi^2}\frac{1}{z^2}$, as is assumed for a Strebel point.
The above differential is essentially Strebel differential for any  value
of $C$ and $\delta$.
It has a pole of residue $m+2$ at $z=\infty$, a zero of valence $m$ at $z=0$, $m+2$
 poles of residue $\sqrt{2}$ and $m+2$ simple zeros.
The length between two points computed using $\vf^{(m)}$ is given by
\be 
l^{(m)}(z_1,z_2)=\left.\frac{i}{\pi}\left(\frac{1}{\sqrt{2}}\ln\left(\frac{\sqrt{2}+\sqrt{1+{\left(\frac{\hat z}{z}\right)}^{2+m}}}
{\sqrt{1+{\left(\frac{\hat z}{z}\right)}^{2+m}}-\sqrt{2}}\right)-\ln\left(z^{\half(m+2)}+\sqrt{{\hat z}^{2+m}+{z}^{2+m}}\right)
\right)\right|_{z_1}^{z_2}\nonumber\\
\ee The critical curve of this differential for two different values of $m$ is depicted in figure \ref{strebFace}.
The picture is that as we approach a Strebel point in the moduli space at each face of the diagram a sphere containing a diagram of the form depicted in
figure \ref{strebFace} pinches off and we are left with a Strebel differential, $\vf_D$, on the sphere. 

Note that the $X$ \eqref{IX} and the $Y$ \eqref{IY} path integrals were non vanishing because of the special structure of poles and 
zeros at the Strebel points. If we have additional zeros and poles, as in the above described extension to non Strebel points,
these path integrals will vanish. This, provides a  localization mechanism for the integration over the moduli space 
to the Strebel points. It would be very interesting to figure out whether the heuristic picture of this section can be made more explicit.


\bibliography{matrix}
\bibliographystyle{JHEP}

\end{document}